\documentclass[%
 amsmath,amssymb,
 aps,
pra,
]{revtex4-2}

\usepackage{dcolumn}
\usepackage{bm}

\usepackage[caption=false]{subfig}
\usepackage[usenames, dvipsnames]{color}
\usepackage{float}   
\usepackage{hyperref} 
\hypersetup{colorlinks=true,citecolor=blue,linkcolor=blue,urlcolor=blue} 
\usepackage{graphicx}

\begin{document}


\title{Dressed Energy Levels in Strongly Interacting Atoms}
\author{Seyed Mostafa Moniri$^1$}
\email{s.m.moniri@gmail.com}
\author{Marjan Fani$^2$}
\email{ma\_fani@sbu.ac.ir}
\author{Elnaz Darsheshdar$^3$}
\email{darsheshdare@gmail.com}

\affiliation{
$^1$Faculty of Science, Golpayegan University of Technology, P.O. Box 87717-67498, Golpayegan, Iran}

\affiliation{%
$^2$Laser and Plasma Research Institute, Shahid Beheshti University, Tehran 19839-69411, Iran
}%
\affiliation{%
$^3$Departamento de F\'{i}sica, Universidade Federal de S\~{a}o Carlos,
P.O. Box 676, 13565-905, S\~{a}o Carlos, S\~{a}o Paulo, Brazil
}%




\date{\today}

\begin{abstract}
We investigate the effect of strong interaction in the dressed energy levels of the two level emitters. Strong dipole-dipole interactions give rise to new sidebands in the fluorescence spectrum due to specific couplings among the collective dressed levels which in turn depends on the spatial configuration of atoms. These couplings are the main responsible for the frequencies and variety of sidebands.
We explain the general method for finding the dressed energy levels for a system of any number of strongly coupled atoms and we solve this problem for two different spatial configurations of three coupled two-level emitters. We show that the coupling among dressed levels and consequently energies and number of sidebands in the fluorescence spectrum are different for each configuration. Thus the fluorescence spectrum of strongly interacting atoms contains information about the number and configuration of atoms. 
\end{abstract}

\maketitle


\section{Introduction}

Dressing of atoms with the incident laser field -in both its classical and quantum descriptions-, or with individual photons is a well-known phenomenon in quantum optics theories. Resulted coupled atom-field energy levels describe well the spectral properties of the system. Fluorescence spectrum is one of the simplest and yet most useful case of the light-matter coupling. It describes the emission of a two-level system (TLS) when is driven strongly by the external laser field. One smart physical interpretation of this spectrum describes a TLS dressed by the laser field \cite{Tannoudji_1977,Reynaud_1983}. This leads to new eigenstates $\left| \pm  \right\rangle$ which are a combination of states $\left| \uparrow ,n \right\rangle $ and $\left| \downarrow ,n+1 \right\rangle $ where $\,\left| \downarrow  \right\rangle ,\,\left| \uparrow  \right\rangle $ are the “ground” and “excited” states of the atom and $n$ is the number of photons. A group of these states builds “manifolds of excitation”. In each manifold, eigenstates are split by the Rabi frequency of the driving field, and the successive manifolds separated by ${{\omega }_{L}}$ which is the one photon energy. The transitions between neighbor manifolds explain the main property: a carrier centered at the laser frequency, and two sidebands symmetrically shifted away by the Rabi frequency of the driving field \cite{Carmichael_1976,Scully_1997,Tannoudji_1998}. 

All mentioned studies so far, are for the single atom effects. In this context, the coupling of the emitters, gives access to new control parameter, as new resonances rise from the interaction. Including the interaction between atoms \cite{Senitzky,Agarwall,Aryeh} new sidebands appear at twice the Rabi frequency from the carrier. Recently it has been shown that these extra sidebands, called the ``baby Mollow", should be observable in a large cloud of cold atoms interacting by the light-mediated dipole-dipole interaction, provided the optical thickness is large enough \cite{Pucci}. 
Spectral properties have been studied in a two atoms system when atoms are weakly interacting by light mediated dipole-dipole interaction, but only one atom is driven by external field. In this case the fluorescence spectrum will not be affected by the interaction and additional sidebands would not appear. Dressed energy levels are still single atom ones \cite{Peng_2019}. 
In the limit of strong dipole-dipole interaction the appearance of the collective dressed energy levels has been shown and photon correlations have been studied in a two atoms system \cite{Darsheshdar2020}.
Although the appearance of collective effects have been reported in the last mentioned work, a description for the collective dressed energy levels for a system of many atoms with the strong interaction have never been done.

In this paper we will explain the general method for obtaining dressed energy levels for systems containing $N$ number of strongly interacting atoms, and the reason of higher number of sidebands in their fluorescence spectrum by solving an example of three atoms system. We also investigate the effect of atomic configuration in the dressed energy levels of strongly interacting atoms and consequently the fluorescence spectrum. For this goal we consider an example of two different spatial configurations of three two-levels atoms and will study the properties of the fluorescence spectrum in this system. We will show that coupling between dressed levels and the additional sidebands will change by changing the atomic spatial configuration. The example of three atoms can be extended to a desired $N$ number of atoms with the cost of computation for diagonalizing the Hamiltonian which is an $N$ by $N$ matrix.  For high number of $N$ it will be an intractable task and approximated methods will be useful.


\section{Fluorescence Spectrum}

Dynamical studies of a system of two-level systems(TLS) driven by an external laser beam and undergoing cooperative emission due to long range interaction mediated by light, can be done by a master equation approach. More specifically we consider identical two-level system (TLS), each of them being described by the spin half angular momentum algebra, at positions ${{\textbf{r}}_{i}}$ and with $\Gamma $ the decay rate of the excited state. Experimentally, this can be either two atoms, two molecules or two quantum dots. The corresponding master equation in the rotating frame of the laser, in the Born, Markov, and rotating-wave approximations is then given by \cite{Agarwal,Das_2008} (we set $\hbar =1$ along the paper) 

\begin{widetext}
\begin{align}
\label{mast}
\frac{\partial \rho }{\partial t}=i\left[ \rho ,{{H}_{\sigma }} \right]+{\mathcal{L}_{\sigma }}\rho,
\end{align}
\begin{align}
\label{Ham}
{{H}_{\sigma }}=\sum\limits_{i}{{H}_{\sigma i}}+\sum\limits_{i,j\ne i}{{H}_{\sigma ij}}=\frac{{{\Omega }}}{2}\sum\limits_{i}({{\operatorname{e}}^{-i\textbf{k}.{{\textbf{r}}_{i}}}}\sigma _{i}^{-}+{{\operatorname{e}}^{i\textbf{k}.{{\textbf{r}}_{i}}}}\sigma _{i}^{+})+\Delta \sum\limits_{i}{\sigma _{i}^{+}\sigma _{i}^{-}}+\Gamma\sum\limits_{i,j\ne i}{{\Omega }_{ij}}\sigma _{i}^{+}\sigma _{j}^{-},
\end{align}
\end{widetext}
where $\rho $ is the density matrix and $\sigma _{i}^{-}$ ($\sigma _{i}^{+}$) is the annihilation (creation) operator of the $i$-th TLS with free energy ${{\omega }_{a}}$. The detuning is the difference between ${{\omega }_{a}}$ and the laser frequency $\Delta ={{\omega }_{a}}-{{\omega }_{L}}$, $\Omega$ is the Rabi frequency of the driving field and $\textbf{k}$ is the wave vector. The Hamiltonian of the system is written in the rotating frame of the laser in Eq. (\ref{Ham}), and the first term is related to atom-field interaction, while the second term corresponds to the atom-atom interaction. The Lindblad super-operator is  
\begin{widetext}
\begin{align}
\label{Lin}
&{\mathcal{L}_{\sigma }}\rho =\frac{\Gamma }{2}\sum\limits_{i}{\mathcal{L}_{\sigma i}}\rho +\frac{\Gamma }{2}\sum\limits_{i,j\ne i}{{\Gamma }_{ij}}{\mathcal{L}_{\sigma ij}}\rho =\frac{\Gamma }{2}\sum\limits_{i}(2\sigma _{i}^{-}\rho \sigma _{i}^{+}-\sigma _{i}^{+}\sigma _{i}^{-}\rho -\rho \sigma _{i}^{+}\sigma _{i}^{-}) \\ \nonumber
& +\frac{\Gamma }{2}\sum\limits_{i,j\ne i}{{\Gamma }_{ij}}(2\sigma _{j}^{-}\rho \sigma _{i}^{+}-\sigma _{i}^{+}\sigma _{j}^{-}\rho -\rho \sigma _{i}^{+}\sigma _{j}^{-}), 
\end{align}
where ${{\Gamma }_{ij}}$ and ${{\Omega }_{ij}}$ are
\begin{align}
\label{gamma}
{{\Gamma }_{ij}}=\frac{3}{2}\left( 1-\cos^{2} {{ \theta_{ij}}} \right)\frac{\sin{(k{r}_{ij})}}{{k{r}_{ij}}}+\frac{3}{2}\left( 1-3\cos^{2} {{\theta_{ij}}} \right)\left( \frac{\cos{(k{r}_{ij})}}{(kr_{ij})^{2}}-\frac{\sin{(k{r}_{ij})}}{(kr_{ij})^{3}} \right),
\end{align}
\begin{align}
\label{omega} 
{{\Omega }_{ij}}=-\frac{3}{4}\left( 1-\cos^{2} {{ \theta_{ij}}} \right)\frac{\cos{(k{r}_{ij})}}{{k{r}_{ij}}}+\frac{3}{4}\left( 1-3\cos^{2} {{\theta_{ij}}} \right)\left( \frac{\sin {(k{r}_{ij})}}{(kr_{ij})^{2}}+\frac{\cos {(k{r}_{ij})}}{(kr_{ij})^{3}} \right),
\end{align}
\end{widetext}
while $\theta_{ij}$ is the angle between the direction of atom dipole moments and the vector joining the $i$th and the $j$th atom, whose distance is ${{r}_{ij}}=\left| {\mathbf{\hat{{r}_{i}}}}-{\mathbf{\hat{{r}_{j}}}} \right|$. The dipole-dipole interaction can be described based on virtual photons exchange between atoms and become the strongest at small interatomic distances. The last term of Eq. (\ref{Lin}), describes the cooperative emission.

We are interested to the one photon spectrum (1PS) of the system which is defined by the Fourier transform of the first-order correlation function of the electric field as bellow
\cite{Pucci} 
\begin{align} 
\label{eq1}
S\left( \omega  \right)=\underset{T\to \infty }{\mathop{\lim }}\,\underset{t\to \infty }{\mathop{\lim }}\,\int_{-T}^{T}{d\tau {{g}^{(1)}}\left( t,\tau  \right){{e}^{-i\omega \tau }}}.
\end{align}
This spectrum is valid in the steady state of the system $t\to \infty $ and the first order correlation function emitted in the direction of the normalized vector $\mathbf{\hat{n}}$
is given by

\begin{align}
\label{eq2} 
{{g}^{(1)}}\left( t,\tau  \right)=\frac{\left\langle E\left( \mathbf{\hat{n}},t \right){{E}^{\dagger }}\left( \mathbf{\hat{n}},t+\tau  \right) \right\rangle }{\left\langle E\left( \mathbf{\hat{n}},t \right){{E}^{\dagger }}\left( \mathbf{\hat{n}},t \right) \right\rangle }, 
\end{align}
with the far field electric field
operator of ${{E}^{\dagger }}\left( \mathbf{\hat{n}},t \right)=\sum\limits_{j=1}^{N}{\sigma _{j}^{-}\left( t \right){{e}^{-ik\mathbf{\hat{n}}.{{\mathbf{r}}_{j}}}}}$. The quantum regression theorem, commonly used for two-time observables is used in calculation of ${{g}^{(1)}}\left( t,\tau  \right)$ \cite{Gardiner_2014,Gisin1993,Brun1996,Breuer1997}. We calculated it numerically using qutip toolbox. We integrated Eq. (\ref{mast}) numerically, for enough long times to obtain the density matrix of the system at the steady state, then we calculated two-time correlation function of ${{g}^{(1)}}\left( t,\tau  \right)$.

\section{Dressed Energy Levels}

In this section we study the collective $N$ atom dressed energy levels. Since this study needs an exact diagonalization of the Hamiltonian which is a large-scale NP problem and intractable due to the exponential growth of the computational costs with the input size, we solve this problem setting $N=3$, the extension to the large scale system is straightforward but very computational resource demanding. We will select two spatial configuration of three atoms located in vertices of an equilateral triangle and in vertices of an isosceles triangle. For each  spatial configuration we will obtain the dressed levels and their transitions related to various sidebands in the 1PS.

\subsection{Equilateral Triangle Spatial Configuration}

Dressed eigenstates of the single atoms that are at resonance with the driving field can be written as $\left| \pm  \right\rangle =\frac{1}{\sqrt{2}}\left( \left| \uparrow ,n-1 \right\rangle \pm \left| \downarrow ,n \right\rangle  \right)$. 
It was shown that turning on the strong interaction in two emitters strongly driven by a laser field leads the collective effects in their dressed energy levels \cite{Darsheshdar2020}.
Increasing the number of atoms up to three when they are located in  vertices of an equilateral triangle cause that the dipole-dipole interaction in  Eq. (\ref{Ham}) generates collective single- or double-excitation eigenstates of
\begin{align}
\label{8}
 & \left| \phi _{1} \right\rangle =a_{1}\left( \left| \uparrow \downarrow \downarrow  \right\rangle +\left| \downarrow \uparrow \downarrow  \right\rangle +\left| \downarrow \downarrow \uparrow \right\rangle  \right) \nonumber \\
 & \left| \phi _{2} \right\rangle =a_{2}\left| \uparrow \downarrow \downarrow  \right\rangle +a_{3}\left( \left| \downarrow \uparrow \downarrow  \right\rangle +\left| \downarrow \downarrow \uparrow  \right\rangle  \right) \nonumber \\ 
 & \left| \phi _{3} \right\rangle =a_{4}\left( \left| \downarrow \uparrow \downarrow  \right\rangle -\left| \downarrow \downarrow \uparrow  \right\rangle  \right) \nonumber \\
   & \left| \phi _{4} \right\rangle =a_{5}\left( \left| \uparrow \uparrow \downarrow  \right\rangle +\left| \uparrow \downarrow \uparrow  \right\rangle +\left| \downarrow  \uparrow \uparrow  \right\rangle  \right) \nonumber \\ 
 & \left| \phi _{5} \right\rangle =a_{6}\left|  \uparrow \uparrow \downarrow \right\rangle +a_{7}\left( \left| \uparrow \downarrow \uparrow  \right\rangle +\left| \downarrow \uparrow \uparrow  \right\rangle  \right) \nonumber \\ 
 & \left| \phi _{6} \right\rangle =a_{8}\left( \left| \uparrow \downarrow \uparrow  \right\rangle -\left| \downarrow \uparrow \uparrow  \right\rangle  \right),
\end{align}

in addition to the ground and three excitation states of

\begin{align}
\label{9}
 & \left| \phi _{0} \right\rangle =\left| \downarrow \downarrow \downarrow  \right\rangle  \nonumber \\  
 & \left| \phi _{7} \right\rangle =\left| \uparrow \uparrow \uparrow  \right\rangle.
\end{align}

Eigenstates introduced in Eqs. (\ref{8}, \ref{9}) can be obtained numerically by diagonalization of the Hamiltonian 
\begin{widetext}
\begin{align}
\label{Ham2}
{{H_{lab}}_{,\sigma }}=\sum\limits_{i}{{H'}_{\sigma i}}+\sum\limits_{i,j\ne i}{{H}_{\sigma ij}}=\sum\limits_{i}{{{\omega }_{a}}\sigma _{i}^{z}}+\Gamma\sum\limits_{i,j\ne i}{{\Omega }_{ij}}\sigma _{i}^{+}\sigma _{j}^{-},
\end{align}
\end{widetext}
in the lab frame where $\sigma_{i}^{z}$ is the energy operator of the \textit{i}th atom. Then $a_i$ in Eqs. (\ref{8}, \ref{9}) are obtained by normalization condition. In this paper we consider the strong atom-atom interaction of $\left| {{\Omega }_{ij}} \right|\gg 1$ and ${{\Gamma }_{ij}}\approx 1$, and strong driving which is defined by the relation $\Omega^2 \gg \Gamma^2+4 |\Gamma\Omega_{ij}|^2$.  In addition to the Rabi frequency of the driving field, this condition can be obtained by tuning the atomic distances $k{{r}_{ij}}$, and the orientation of the atomic dipoles ${{\theta }_{ij}}$, in Eqs. (\ref{omega}) and (\ref{gamma}).

\begin{figure}[H]  
	\centering
	\subfloat[]{
		\includegraphics[width=0.4\textwidth]{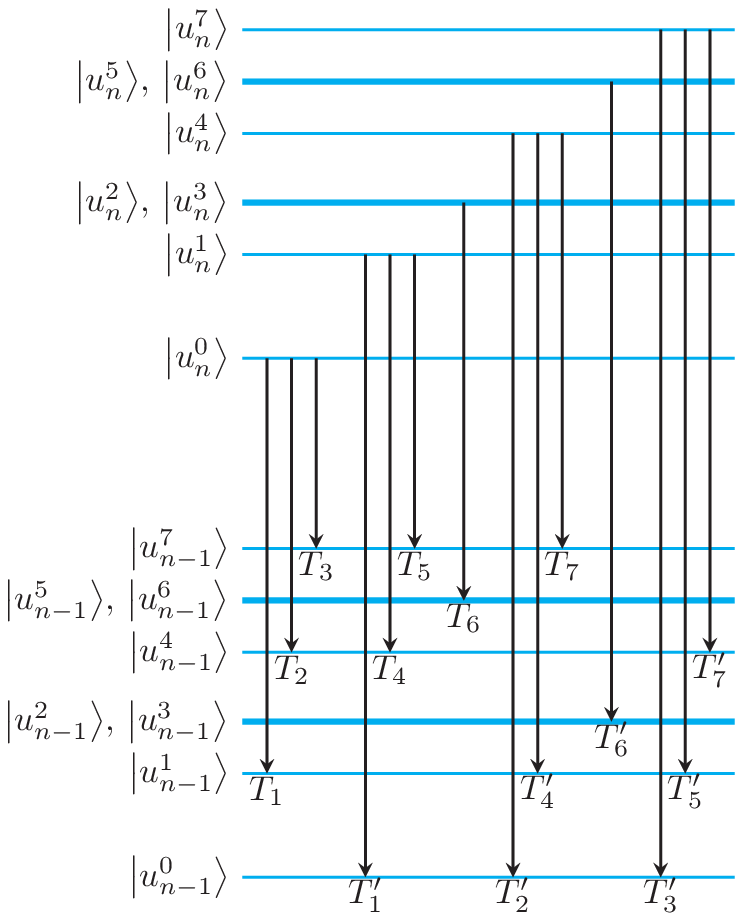}
        \label{fig:1-a}}\\ 
    \subfloat[]{
		\includegraphics[width=0.8\textwidth]{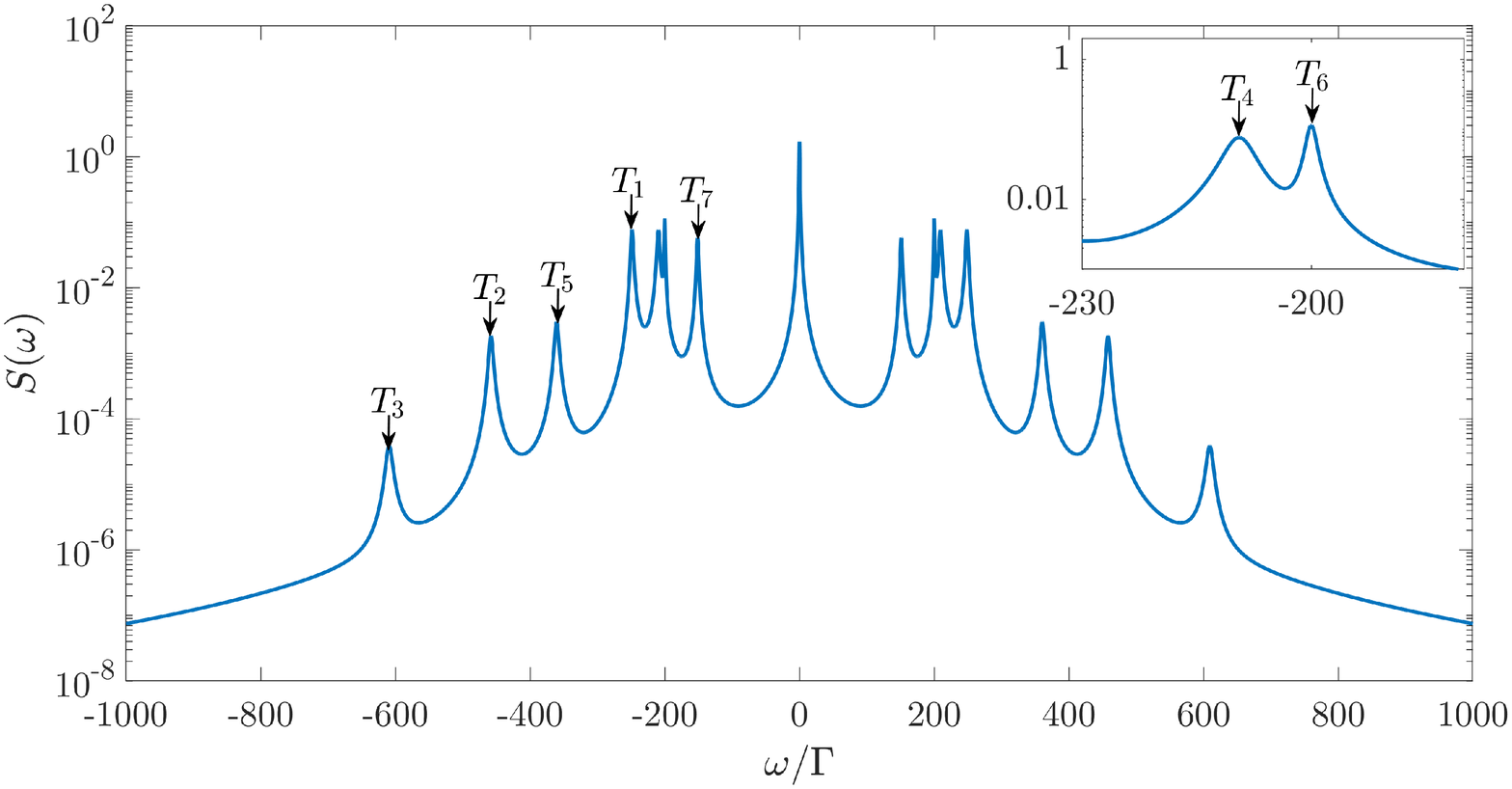}
		\label{fig:1-b}}
	\caption{(a) Collective dressed energy levels whose transitions between manifolds account for the main phenomenology of 1PS, shown in the rotating frame of the laser. (b) Fluorescence spectrum. Both (a) and (b) are related to three strongly interacting atoms in the equilateral triangle spatial configuration. Different peaks are indicated with different $T_i$ signs corresponding to the related transitions in dressed levels. $T'_i$ are not shown in the spectrum as they are symmetrical with the $T_i$ transitions. The inset shows frequency range of $\omega=[-230,-180]$ in order to see it in higher resolution. The results were obtained by driving three strongly interacting TLS resonantly with $\Omega =200\Gamma$, $\Gamma$ which is the decay rate of the individual atoms sets the frequency unit throughout the text. The interaction between atoms is determined by setting $kr=0.01$ and $\theta ={{\cos }^{-1}}(\frac{1}{\sqrt{3}})$ in Eqs. (\ref{Lin}) and (\ref{gamma}). Also ${{\omega }_{L}}$ (laser frequency) , is taken as a reference for the frequency.}
	\label{fig:1}
\end{figure}

We follow the approach of \cite{Compagno} to study collective three atoms dressed levels. We consider the following basis 
\begin{align} 
\label{10}
 & \left|\phi_0^n\right\rangle =\left|\phi_0\right\rangle \left|n\right\rangle   \nonumber \\  
 & \left|\phi_i^n\right\rangle =\left|\phi_i\right\rangle \left|n-1\right\rangle \quad i=1,2,3   \nonumber \\
 & \left|\phi_j^n\right\rangle =\left|\phi_j\right\rangle \left|n-2\right\rangle \quad j=4,5,6  \nonumber \\
 & \left|\phi_7^n\right\rangle =\left|\phi_7\right\rangle \left|n-3\right\rangle.
\end{align}

The above eight-dimensional subspace is characterized by an eigenvalue of the operator ${{N}_{T}}={N_\nu}+\sum\limits_{i}{\frac{\sigma _{i}^{z}}{2}}+\frac{1}{2}$, where ${N_\nu}$ is the photon number. For $N$ particles system the above  subspace is ${{2}^{N}}$-dimensional and the photon number is restored to obtain hybrid atom-field states having the same eigenvalue of ${{N}_{T}}$.
In this new basis, the eigenstates of the Hamiltonian in Eq. (\ref{Ham}), which are ``atoms-light" collective dressed state, reads
\begin{align} 
\label{dressed}
  & \left| u_{0}^{n} \right\rangle =b_{1}\left| \phi _{0}^{n} \right\rangle +b_{2}\left| \phi _{1}^{n} \right\rangle +b_{3}\left| \phi _{4}^{n} \right\rangle +b_{4}\left| \phi _{7}^{n} \right\rangle  \nonumber \\ 
 & \left| u_{1}^{n} \right\rangle =b_{5}\left| \phi _{0}^{n} \right\rangle +b_{6}\left| \phi _{1}^{n} \right\rangle  + b_{7}\left| \phi _{4}^{n} \right\rangle +b_{8}\left| \phi _{7}^{n} \right\rangle  \nonumber \\ 
 & \left| u_{2}^{n} \right\rangle =b_{9}\left| \phi _{2}^{n} \right\rangle +b_{10}\left| \phi _{3}^{n} \right\rangle +b_{11}\left| \phi _{5}^{n} \right\rangle +b_{12}\left| \phi _{6}^{n} \right\rangle \nonumber \\ 
 & \left| u_{3}^{n} \right\rangle =b_{13}\left| \phi _{2}^{n} \right\rangle +b_{14}\left| \phi _{3}^{n} \right\rangle +b_{15}\left| \phi _{5}^{n} \right\rangle +b_{16}\left| \phi _{6}^{n} \right\rangle \nonumber \\
 & \left| u_{4}^{n} \right\rangle =b_{17}\left| \phi _{0}^{n} \right\rangle +b_{18}\left| \phi _{1}^{n} \right\rangle +b_{19}\left| \phi _{4}^{n} \right\rangle +b_{20}\left| \phi _{7}^{n} \right\rangle \nonumber \\ 
 & \left| u_{5}^{n} \right\rangle =b_{21}\left| \phi _{2}^{n} \right\rangle +b_{22}\left| \phi _{3}^{n} \right\rangle +b_{23}\left| \phi _{5}^{n} \right\rangle +b_{24}\left| \phi _{6}^{n} \right\rangle \nonumber \\ 
 & \left| u_{6}^{n} \right\rangle =b_{25}\left| \phi _{2}^{n} \right\rangle +b_{26}\left| \phi _{3}^{n} \right\rangle +b_{27}\left| \phi _{5}^{n} \right\rangle +b_{28}\left| \phi _{6}^{n} \right\rangle  \nonumber \\ 
 & \left| u_{7}^{n} \right\rangle =b_{29}\left| \phi _{0}^{n} \right\rangle +b_{30}\left| \phi _{1}^{n} \right\rangle +b_{31}\left| \phi _{4}^{n} \right\rangle +b_{32}\left| \phi _{7}^{n} \right\rangle, 
\end{align}
 where $\left| u_{i}^{n} \right\rangle$ and $b_{i}$ are obtained by numerical diagonalization of the Hamiltonian (\ref{Ham}) and then writing its eigenstates in the basis of \ref{10}, in addition to the normalization condition. 

The ``atoms-light" collective dressed states of this configuration is thus an octet of $\left| u_{n}^{i} \right\rangle$, with $i=0,1,..,7$ and numerically obtained eigenenergies $E_{n}^{i}$. Dressed states of each manifold are non-equally spaced by 

\begin{equation}
    \Delta_{ij}=-\Delta_{ji}\equiv E_{n}^{i}-E_{n}^{j},
\end{equation}
and the energy difference between two neighboring manifolds is equal to ${{\omega }_{L}}$. Manifolds of collective dressed states of three strongly interacting atoms in equilateral triangle configuration are shown in Fig. (\ref{fig:1-a}).

The spectral shape of 1PS can be obtained by solving the master equation of Eq. (\ref{mast}), and this spectrum is understood as transitions between the dressed levels. Similar to the one atom dressed levels, transitions from one level of higher energy manifold (with $E_{n}^{i}$) to the same energy level in the lower energy one (with $E_{n-1}^{i}$), leads to the central peak. It leaves the state of the three TLS unchanged ($\left| u_{n}^{i} \right\rangle \to \left| u_{n-1}^{i} \right\rangle $ for $i=0,1,..,7$), and the transition frequency is ${{\omega }_{L}}$. Other various transitions and the related peaks are shown in Figs. (\ref{fig:1-a}) and (\ref{fig:1-b}). All transitions are hereafter given in the laser frame.

Dressed energy levels are characterized by entanglement between atomic and field states, and various transitions between these levels that are matched with the peaks of the 1PS will be explained based on these couplings. For instance as it can be seen from the Eqs. (\ref{dressed}) $ \left| u_{n}^{2} \right\rangle$ and $ \left| u_{n}^{3} \right\rangle$  are coupled only with $ \left| u_{n}^{5} \right\rangle$ and $ \left| u_{n}^{6} \right\rangle$ thus they have transitions only between each other in the successive manifolds. Note that  $ \left| u_{n}^{2} \right\rangle$ and $ \left| u_{n}^{3} \right\rangle$ are degenerate, the same holds true for the  $ \left| u_{n}^{5} \right\rangle$ and $ \left| u_{n}^{6} \right\rangle$. Other states that are not coupled to these four energy levels do not have transitions to (or from) them. 
Consequently it can be seen from Fig. (\ref{fig:1}) that $ \left| u_{n}^{i} \right\rangle$ with $i=0,1,4,7$ transit only between each other. These couplings are the main responsible for the fourteen number of peaks in this configuration.
In the following of the text we will study the couplings and sidebands of the fluorescence spectrum when atoms are located in the vertices of an isosceles triangle.  

The fluorescence spectrum of Fig. (\ref{fig:1-b}) indicates pure collective (configuration induced) effects by the presence of fourteen sidebands, with frequency differences compared to the central peak of $\pm \Delta_{ij}$, in this configuration of atoms.

\subsection{Isosceles Triangle Spatial Configuration}

When three atoms located at the vertices of an isosceles triangle, the collective single- and double -excitation eigenstates, together with the ground and three exited states are:

\begin{align}
\label{12+1}
& \left| \phi _{0} \right\rangle =\left| \downarrow \downarrow \downarrow  \right\rangle  \nonumber\\ 
 & \left| \phi _{1} \right\rangle =a_1\left| \uparrow \downarrow \downarrow  \right\rangle +a_2\left( \left| \downarrow \uparrow \downarrow  \right\rangle +\left| \downarrow \downarrow \uparrow  \right\rangle  \right) \nonumber\\ 
 & \left| \phi _{2} \right\rangle =a_3\left( \left| \downarrow \uparrow \downarrow  \right\rangle -\left| \downarrow \downarrow \uparrow  \right\rangle  \right) \nonumber\\ 
 & \left| \phi _{3} \right\rangle =a_4\left| \uparrow \downarrow \downarrow  \right\rangle +a_5\left( \left| \downarrow \uparrow \downarrow  \right\rangle +\left| \downarrow \downarrow \uparrow  \right\rangle  \right)\nonumber\\
  & \left| \phi _{4} \right\rangle =a_6\left| \downarrow \uparrow \uparrow  \right\rangle +a_7\left( \left| \uparrow \uparrow \downarrow  \right\rangle +\left| \uparrow \downarrow \uparrow  \right\rangle  \right) \nonumber\\ 
 & \left| \phi _{5} \right\rangle =a_8\left( \left| \uparrow \uparrow \downarrow  \right\rangle -\left| \uparrow \downarrow \uparrow  \right\rangle  \right) \nonumber\\ 
 & \left| \phi _{6} \right\rangle =a_9\left| \downarrow \uparrow \uparrow   \right\rangle +a_{10}\left( \left| \uparrow \uparrow \downarrow  \right\rangle +\left| \uparrow \downarrow \uparrow  \right\rangle  \right) \nonumber\\
 & \left| \phi _{7} \right\rangle =\left| \uparrow \uparrow \uparrow  \right\rangle.
\end{align}

Since the configuration of atoms is different from the equilateral triangle, the basis of (\ref{12+1}) which is equivalent to the basis of (\ref{8},\ref{9}), i.e. the eigenstates of the (\ref{Ham2}) in the lab frame has different couplings.
In this structure of the atoms, since the $\left| \phi _{i} \right\rangle$ have the same atomic excitation compared to the equilateral triangle configuration, we can again consider the basis of Eq.(\ref{10})  and obtain eigenstates of the Hamiltonian of (\ref{Ham}) in this basis. 

The eigenstates of the atom-light system are composed by the collective dressed states incorporating the eigenstates of  Hamiltonian \eqref{Ham} for atoms with light-mediated dipole-dipole interactions, and the photon number states of the light field i.e., the $n$-excitation manifold for our system is given by
 
\begin{align}
\label{14}
 \left| u_{0}^{n} \right\rangle =&d_{1}\left| \phi _{0}^{n} \right\rangle +d_{2}\left| \phi _{1}^{n} \right\rangle +d_{3}\left| \phi _{3}^{n} \right\rangle +d_{4}\left| \phi _{4}^{n} \right\rangle \nonumber\\ 
 &+d_{5}\left| \phi _{6}^{n} \right\rangle +d_{6}\left| \phi _{7}^{n} \right\rangle  \nonumber\\ 
 \left| u_{1}^{n} \right\rangle =&d_{7}\left| \phi _{0}^{n} \right\rangle +d_{8}\left| \phi _{1}^{n} \right\rangle +d_{9}\left| \phi _{3}^{n} \right\rangle +d_{10}\left| \phi _{4}^{n} \right\rangle \nonumber\\ 
 &+d_{11}\left| \phi _{6}^{n} \right\rangle +d_{12}\left| \phi _{7}^{n} \right\rangle  \nonumber\\ 
 \left| u_{2}^{n} \right\rangle =&d_{13}\left| \phi _{2}^{n} \right\rangle +d_{14}\left| \phi _{5}^{n} \right\rangle  \nonumber\\ 
 \left| u_{3}^{n} \right\rangle =&d_{15}\left| \phi _{0}^{n} \right\rangle +d_{16}\left| \phi _{1}^{n} \right\rangle +d_{17}\left| \phi _{3}^{n} \right\rangle +d_{18}\left| \phi _{4}^{n} \right\rangle \nonumber\\ 
 &+d_{19}\left| \phi _{6}^{n} \right\rangle +d_{20}\left| \phi _{7}^{n} \right\rangle  \nonumber\\ 
 \left| u_{4}^{n} \right\rangle =&d_{21}\left| \phi _{0}^{n} \right\rangle +d_{22}\left| \phi _{1}^{n} \right\rangle +d_{23}\left| \phi _{3}^{n} \right\rangle +d_{24}\left| \phi _{4}^{n} \right\rangle \nonumber\\ 
 &+d_{25}\left| \phi _{6}^{n} \right\rangle +d_{26}\left| \phi _{7}^{n} \right\rangle  \nonumber\\ 
 \left| u_{5}^{n} \right\rangle =&d_{27}\left| \phi _{2}^{n} \right\rangle +d_{28}\left| \phi _{5}^{n} \right\rangle  \nonumber\\ 
 \left| u_{6}^{n} \right\rangle =&d_{29}\left| \phi _{0}^{n} \right\rangle +d_{30}\left| \phi _{1}^{n} \right\rangle +d_{31}\left| \phi _{3}^{n} \right\rangle +d_{32}\left| \phi _{4}^{n} \right\rangle \nonumber\\ 
 &+d_{33}\left| \phi _{6}^{n} \right\rangle +d_{34}\left| \phi _{7}^{n} \right\rangle  \nonumber\\ 
 \left| u_{7}^{n} \right\rangle =&d_{35}\left| \phi _{0}^{n} \right\rangle +d_{36}\left| \phi _{1}^{n} \right\rangle +d_{37}\left| \phi _{3}^{n} \right\rangle +d_{38}\left| \phi _{4}^{n} \right\rangle \nonumber\\ 
 &+d_{39}\left| \phi _{6}^{n} \right\rangle +d_{40}\left| \phi _{7}^{n} \right\rangle.
\end{align}

\begin{figure}[H]  
	\centering
	\subfloat[]{
		\includegraphics[width=0.4\textwidth]{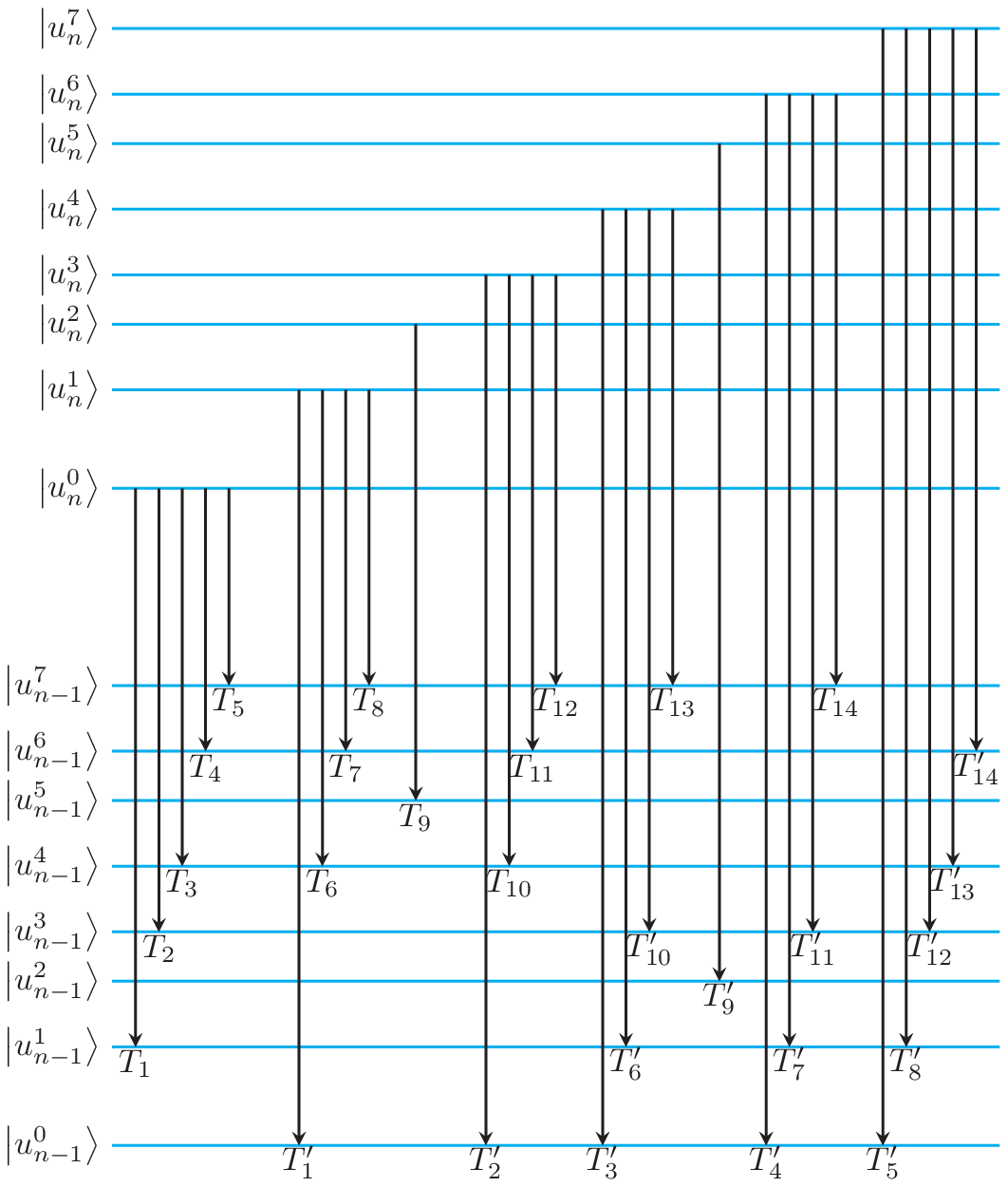}
        \label{fig:2-a}}\\ 
    \subfloat[]{
		\includegraphics[width=0.8\textwidth]{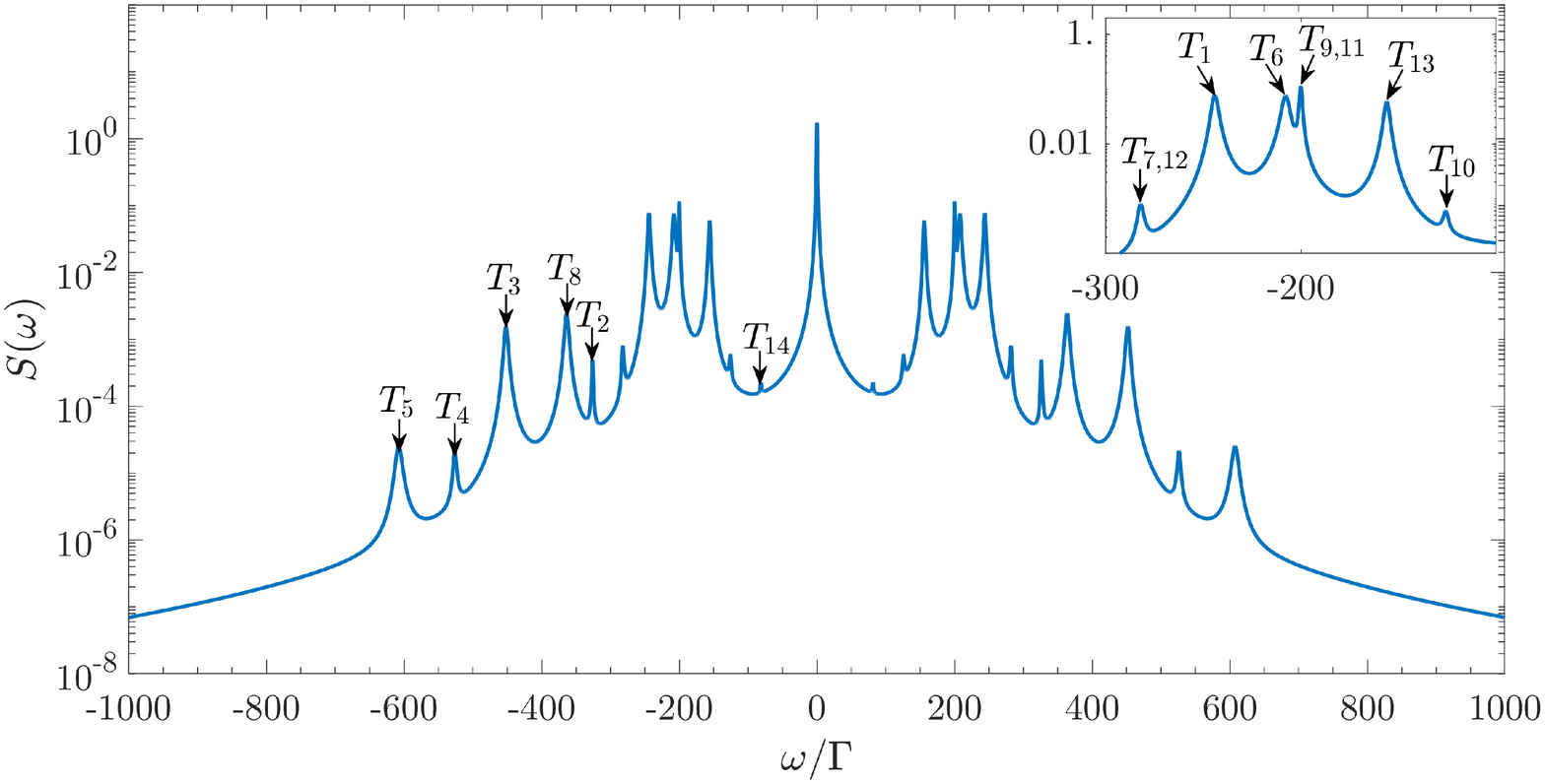}
		\label{fig:2-b}}
	\caption{(a) Collective dressed energy levels whose transitions between manifolds account for the main phenomenology of 1PS, shown in the rotating frame of the laser. (b) Fluorescence spectrum. Both (a) and (b) are related to three strongly interacting atoms in the isosceles triangle spatial configuration. Different peaks are indicated with different $T_i$ signs corresponding to the related transitions in dressed levels. $T'_i$ are not shown in the spectrum as they are symmetrical with the $T_i$ transitions. Note that the transitions of $T_7$ and $T_{12}$ are degenerate, the same holds true for the $T_9$ and $T_{11}$. The inset shows frequency range of $\omega=[-300,-100]$ in order to see it in higher resolution. The results were obtained by driving three strongly interacting TLS resonantly with $\Omega =200\Gamma$, $\Gamma$ which is the decay rate of the individual atoms sets the frequency unit throughout the text. The interaction between atoms is determined by setting $kr=0.01$ for the two equal sides of the triangle when the angle between them is $\pi/2$ and $\theta ={{\cos }^{-1}}(\frac{1}{\sqrt{3}})$ in Eqs. (\ref{Lin}) and (\ref{gamma}). Also ${{\omega }_{L}}$ (laser frequency) , is taken as a reference for the frequency.}
	\label{fig:2}
\end{figure}

As it can be seen from Eq. (\ref{14}), $ \left| u_{2}^{n} \right\rangle$ and $\left| u_{5}^{n} \right\rangle$ are coupled only with each other thus their transitions in successive manifolds are limited only between themselves. As a result other dressed levels of $\left| u_{i}^{n} \right\rangle$, $i=0,1,3,4,6,7$ have transitions only between each other since they do not coupled to $ \left| u_{2}^{n} \right\rangle$ and $\left| u_{5}^{n} \right\rangle$. This can be understood well by checking the energy of the peaks in 1PS and compare them with the (numerically obtained) transition energies between various dressed levels. Transitions between dressed energy levels and the related peaks in the 1PS are shown in Figs. (\ref{fig:2}).

Having the same atomic excitation for two states leads to the same value for the $N_\nu$ -the photon number operator- as well, thus transitions between them in successive manifolds may lead to a peak at $\omega_L$ which is set at the origin of the 1PS plots. One example of these transitions is $\left| u_{4}^{n} \right\rangle \to \left| u_{6}^{n-1} \right\rangle$. Checking carefully the sidebands energies of the 1PS, one can see that there is no peak matched with the transition energy of these two levels in the successive manifolds and since in this transition only the photon excitation number $n$ will be decreased, thus it is equivalent to the $ \left| u_{i}^{n} \right\rangle \to \left| u_{i}^{n-1} \right\rangle$. Note that in the equilateral triangle configuration $\left| u_{4}^{n} \right\rangle$ and $\left| u_{6}^{n} \right\rangle$ are not coupled to each other and we didn't see this effect.

The transformation into a new couplings between dressed states, due to the new configuration of dipole-dipole interactions, leads to twenty-four sidebands in the 1PS as shown in Fig. (\ref{fig:2-b}).  These sidebands are collective and configuration induced, corresponding to twenty-four resonant frequencies $\pm {{\Delta }_{ij}}$ not presented for equilateral triangle.

\section{Conclusion and perspectives}

Strong interactions between three two-level emitters give rise to a series of new sidebands in the fluorescence spectrum, whose shifts from the atomic transition and their number depend on the couplings among dressed levels which in turn depend on spatial configuration of atoms, in addition to the interaction strength and the driving field.
Similarly to the three emitters, for a system of many particles the appearance of more additional sidebands is expected. This can be shown by exact diagonalization of the Hamiltonian which is a large-scale NP problem and intractable due to the exponential growth of the computational costs with the input size.
To circumvent this problem, one needs to deal with approximated methods, for instance the Bogoliubov-Born-Green-Kirkwood-Yvon (BBGKY) hierarchy \cite{Pucci}.
Benchmarking of this approximation against exact results was discussed in \cite{Pucci} only for the limit of weak interaction explained in \cite{Darsheshdar2020} which occurs when  the  distance  between  the  emitters is comparable or larger than the optical wavelength. In this regime the BBGKY hierarchy is in very good agreement with the exact results obtained by the “Quantum Toolbox in Python” \cite{Pucci}.

Besides the presented results in this paper, we also have monitored the BBGKY hierarchy results (without a discrete phase space
representation of atoms \cite{Wootters_1987,Schachenmayer_2015} which is used in \cite{Pucci,Pucci_2016}), and compared them with the exact solution in the strong interaction regime.
We used BBGKY hierarchy considering up to third order of correlation which is the best approximation for four particles and suitable for the quantity of \ref{eq2} as it is a two time two particle correlation function. We neglect the 3-operators terms of $\left\langle \sigma _{i}^{\alpha }\sigma _{j}^{\beta }\sigma _{k}^{\gamma } \right\rangle $  with $i\ne j, k$.
Furthermore, we neglect the 4-operator terms, since they involve 3-operators with $i\ne j\ne k$.

For a chain of four atoms with $kr=0.01$, $\theta=\cos^{-1}(1/\sqrt{3})$, $\Omega=200$, the above explained BBGKY hierarchy didn't give the correct number and energies of the sidebands, as compared to the exact solution (not shown here). The same holds true for five atoms at the same spatial configuration and interaction parameters. This confirms that despite the method is highly accurate for large scale systems with weak long-range interactions, the accuracy is not happening in the strong interaction regime. Note that based on the transparency-induced dark states~ \cite{Agarwal2006,Cho2007,Yavuz2007,Gorshkov2008}, which allows for the generation of subwavelength cold atom structures~\cite{Miles2013,Wang2018,Subhankar2019,Tsui2020}, some designs have been suggested to surpass limitations of manipulating atoms in scales below the diffraction limit.
It is needed to mention that the strong interaction regime can be obtained by tuning both the atomic distances and also the angle between atoms dipole moments and vector joining them.

\section*{Acknowledgment}

 E.\,D. benefited from Grant from S\~ao Paulo Research Foundation (FAPESP, Grant Number 2018/10813-2.

\clearpage
\bibliographystyle{apsrev4-2}
\bibliography{Refsn}

\end{document}